\documentclass[twocolumn,prl]{revtex4}

\usepackage{graphicx}

\begin{document}

\title{Increasing the output of a Littman-type laser by use of an intracavity {F}araday rotator}

\author{Rebecca Merrill}
\author{Rebecca Olson}
\author{Scott Bergeson}
\author{Dallin S. Durfee}
 \affiliation{Department of Physics and Astronomy, Brigham Young University,
Provo, Utah 84602, USA}

\date{\today}

\begin{abstract}

We present a new method of external-cavity diode laser grating
stabilization which combines the high output power of the Littrow
design with the fixed output pointing of the Littman-Metcalf
design. Our new approach utilizes a Faraday-effect optical
isolator inside the external cavity.  Experimental testing and a
model which describes the tuning range and optimal tuning
parameters of the laser are described. Preliminary testing of this
design has resulted in a short-term linewidth of 360 kHz and a
side-mode suppression of 37 dB.  The laser tunes mode-hop free
over 7 GHz and we predict that much larger tuning ranges are
possible.  Published in Applied Optics, Vol. 43, No. 19.
\copyright 2004 Optical Society of America.

\end{abstract}


\maketitle

\section{Introduction}

Inexpensive single-mode laser diodes are readily available at a
variety of wavelengths from the red to the near-infrared. They
require no maintenance, consume little electrical power, require
almost no cooling, can have very high amplitude and pointing
stability, and can be easily modulated at high frequencies (see
\cite{Wieman91ud} and the references therein). Using optical
feedback techniques and employing stable current and temperature
controllers \cite{Libbrecht93al, Bradley90if, Wieman91ud}, laser
diodes can be made to operate at a single frequency with a narrow
linewidth, making them suitable for applications such as
precision spectroscopy and laser cooling. Stabilized diode
systems can often replace considerably more expensive systems
requiring significant infrastructure and regular maintenance.

The two commonly used diode stabilization schemes, the Littrow
\cite{Hansch72rp} and Littman-Metcalf \cite{Littman78sn,
Shoshan77no} designs, each have their advantages. The simplest of
the two designs is the Littrow scheme.  In this arrangement a
reflection grating is placed in front of a collimated diode at an
angle such that the first order diffraction peak at a particular
wavelength is directed back into the diode.  Mode competition
then favors this wavelength. The zeroth-order grating reflection
is used as the output-coupler to extract light from the cavity.
Light only diffracts off of the grating once per round trip
through the cavity in this configuration. As discussed below,
this can result in higher output powers than is possible with the
Littman design. This can be of great importance due to the low
power typical of single-mode diode laser systems relative to what
is possible with other technologies. While it is possible to
amplify a weak laser beam or use a weak stabilized beam to
injection-lock a free running diode, this adds cost and
complexity.

The main drawback of the Littrow design is that as the laser is
tuned by rotating the grating, the pointing of the zeroth-order
output beam changes.  This is not the case in the Littman-Metcalf
design.  In a Littman laser the grating is placed in front of the
diode at an angle closer to grazing incidence, such that the
diffracted light does not return to the laser diode. Instead, the
diffracted beam is directed to a mirror. Depending on the angle of
the mirror, a particular wavelength will be precisely
retro-reflected back to the grating such that it returns to the
diode after diffracting a second time. Like the Littrow design,
the zeroth-order beam from the grating is used to couple light
out of the cavity.  The laser can be tuned with the mirror while
keeping the grating fixed such that the output beam pointing does
not change as the laser is tuned.

Because the angle between the incident beam and the grating is
not fixed to the Littrow angle, it is possible to adjust the
cavity of a Littman laser to accommodate grating angles closer or
further from grazing incidence, allowing the diffraction
efficiency of the grating to be ``tuned'' to produce the minimum
necessary feedback, thereby optimizing the intensity of the
output beam. Another advantage of the Littman design is that
mode-hop free tuning across the entire gain curve of the diode
can be accomplished by simply pivoting the tuning mirror about a
fixed axis \cite{Liu81ng, McNicholl85sc}.

The disadvantage of the Littman design is its inherently lower
power.  In the Littman scheme a single round-trip through the
cavity involves diffracting twice off of the grating.  This has
the fortunate side-effect of increased side-mode suppression. But
the double diffraction means that the grating efficiency needs to
be larger in order for sufficient light to be coupled back to the
diode. This results in less light being coupled out in the
zeroth-order of the first grating pass.  The ``missing power'' is
coupled out in the zeroth-order beam of the second grating pass
in a secondary output beam which does not remain fixed as the
laser is tuned.  As such, commercial Littman-configuration lasers
typically produce just over half of the power of comparable
Littrow-configuration devices \cite{sacher13}.

\begin{figure}
\includegraphics[height=8.2 cm, angle=-90]{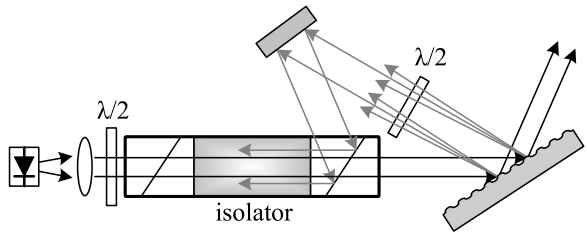}

\caption{ Schematic of the grating stabilization scheme. The
laser is first collimated, and then the polarization is rotated
with a $\lambda / 2 $ plate such that all of the light passes
through the Faraday-effect isolator. Upon exiting the isolator,
the light strikes a diffraction grating. The zeroth-order
specular reflection is used as the output-coupler for the laser.
The first-order diffracted light passes through a $\lambda / 2 $
plate which rotates the polarization by 90 degrees, and is then
reflected by a mirror into one of the rejection ports of the
isolator.  The frequency of light which is coupled back into the
laser is determined by the angle of the grating and the position
of the mirror. \label{fig:design}}
\end{figure}

We have devised and tested a new external-cavity grating
stabilization scheme which combines the single-diffraction power
advantage of the Littrow configuration with the
frequency-independent output pointing and freedom of grating
alignment of the Littman-Metcalf scheme.  Like the Littman design,
in our scheme a mirror, rather than the grating, is used to tune
the laser, keeping the output beam pointing fixed.  But rather
than reflecting the light back at the grating, in this new scheme
the mirror directs the light into a rejection port of an
intra-cavity Faraday-effect optical isolator such that the light
is directly coupled back into the laser diode without striking the
grating a second time. This design is similar to experiments in
which a slave laser has been injection locked by coupling a master
laser through the rejection port of an isolator \cite{Bouyer96ms}.
In this case, however, the laser is ``injection locked'' to
itself. The design is illustrated in Fig.\ \ref{fig:design}.

We should point out that our technique is not the only way to
improve output power while keeping beam pointing fixed.  For
example, various multi-arm grazing incidence cavities developed to
increase the mode selectivity \cite{Binks98lm} or lower the
threshold \cite{Zhang97lt} of Littman-type dye lasers could be
used to increase the output of a Littman diode lasers.  Other
previously demonstrated techniques include the employment of a
mirror moving in conjunction with the grating to correct the beam
pointing of a Littrow laser at the cost of a small parallel
displacement of the beam \cite{Hawthorn01lc}.  Our technique has
the advantage that it requires only one moving element and creates
no output beam displacement. And, unlike the multi-arm cavities in
which extra boundary conditions must be met in order to avoid
mode-hopping, this new the scheme uses a simpler single loop
cavity.

We should also note that due to the size of the isolator, there
are practical limits on how small the external cavity can be in
our design, possibly causing longitudinal modes to be stacked
closer together than would be desirable. Nevertheless, despite
the slightly larger-than-average cavity length in our setup we
have not had difficulty keeping the laser running in a single
longitudinal mode. This limit could be mitigated by using a
miniature isolator at the expense of higher isolator losses.

\section{Experiment}

For our first test of this new stabilization scheme we utilized a
657 nm ``Circu-Laser'' diode from Blue Sky Research
\cite{circulaser} collimated with an aspheric lens to a Gaussian
beam waist radius of 0.5 mm. This diode, originally purchased for
a different purpose, was not an optimum choice for this work
because it lacked an anti-reflection (AR) coating on its front
facet.  Despite the greater susceptibility of an un-coated diode
to mode-hopping \cite{Yan92mo}, we have achieved excellent results
with this laser. Using a Fabry-Perot spectrum analyzer we have
verified a mode-hop free tuning range of 7 GHz. Theoretical
calculations presented in the last half of this paper suggest that
much larger tuning ranges are possible.

In our current implementation the diode laser is placed 5.7 cm
from the end of a 11.8 cm long isolator.  Prior to entering the
isolator the laser is collimated with a 1 mm focal length
aspheric lens and passed through a half-wave plate to align the
polarization of the beam with the input polarizer of the
isolator.  A holographic grating is placed 2.8 cm from the
isolator's output polarizer.  The tuning mirror is mounted to a
three-axis piezo-electric kinematic mount.  The laser is tuned
coarsely by manually adjusting threaded actuators on the mount,
and fine tuning is done using the piezos.  By scanning the
voltages applied to the piezos such that the voltages on each
side of the mount differ only by a proportionality constant, the
mirror can be made to both rotate and translate as the laser is
scanned, effectively causing the mirror to pivot about an axis
offset from the center of the mirror.

Using our calculations, we found that the optimum pivot point for
our configuration is about 17 cm from the center of the mirror
(see Eq.\ \ref{eq:pivot} in Sec.\ \ref{sec:theory}). This
relatively large length means that the mirror must be translated a
considerable distance per degree of rotation about its center to
achieve the optimum tuning range. As a result, scanning the laser
frequency over 7 GHz required the piezos to be scanned over their
entire voltage range. Scans longer than 7 GHz could possibly be
accomplished by changing the dimensions of our cavity to move the
optimum pivot point closer to the mirror or by using actuators
with a greater range of motion.

The spectral properties of our laser are typical of what would be
expected for a Littrow configuration. Using Fabry-Perot spectrum
analyzers we have measured a short-term linewidth of 360 kHz and
a side-mode suppression of 37 dB. Due to losses in the optical
isolator, the output power is somewhat less than in a comparable
Littrow laser. But since most installations of grating stabilized
lasers require an isolator on the output of the laser, this is
not a serious disadvantage.  Due to the low finesse typical of
the external cavity of grating stabilized lasers, the loss due to
the isolator inside the cavity is comparable to the loss that
would be caused by an isolator external to the cavity. The
intra-cavity isolator in our design provides the same immunity to
reflections as an external isolator.

\section{Mode-Hop Free Tuning Theory \label{sec:theory}}

In order to keep the laser from jumping between longitudinal modes
as the frequency of the laser is scanned, the length of the cavity
must increase in proportion to the wavelength of light injected
back into the diode.  If $\lambda_0$ represents the wavelength of
light coupled back into the diode and $S_0$ represents the
round-trip optical path length of the cavity before the laser is
tuned, then mode-hop free tuning is achieved when
\begin{equation}
\label{eq:modehopcondition} \frac{\Delta S}{S_0} = \frac{\Delta
\lambda}{\lambda_0}
\end{equation}
where $\Delta S$ and $\Delta \lambda$ represent the shift in the
cavity length and the injected wavelength from their nominal
values.  This condition ensures that as the wavelength of the
laser is tuned, the round-trip length of the cavity is always a
fixed integer times the wavelength of the laser:
\begin{equation}
    S = m \lambda \label{eq:nnodes}
\end{equation}
Due to the low finesse of typical grating cavities, grating
stabilized lasers can operate in a single mode even when $m$ is
not precisely an integer.  But if $m$ increases or decreases by
more than 0.5, the losses in the current mode become greater than
the losses in an adjacent mode and with near certainty the laser
will hop to the next mode.

\begin{figure}
\includegraphics[width=8.1cm in, angle=0]{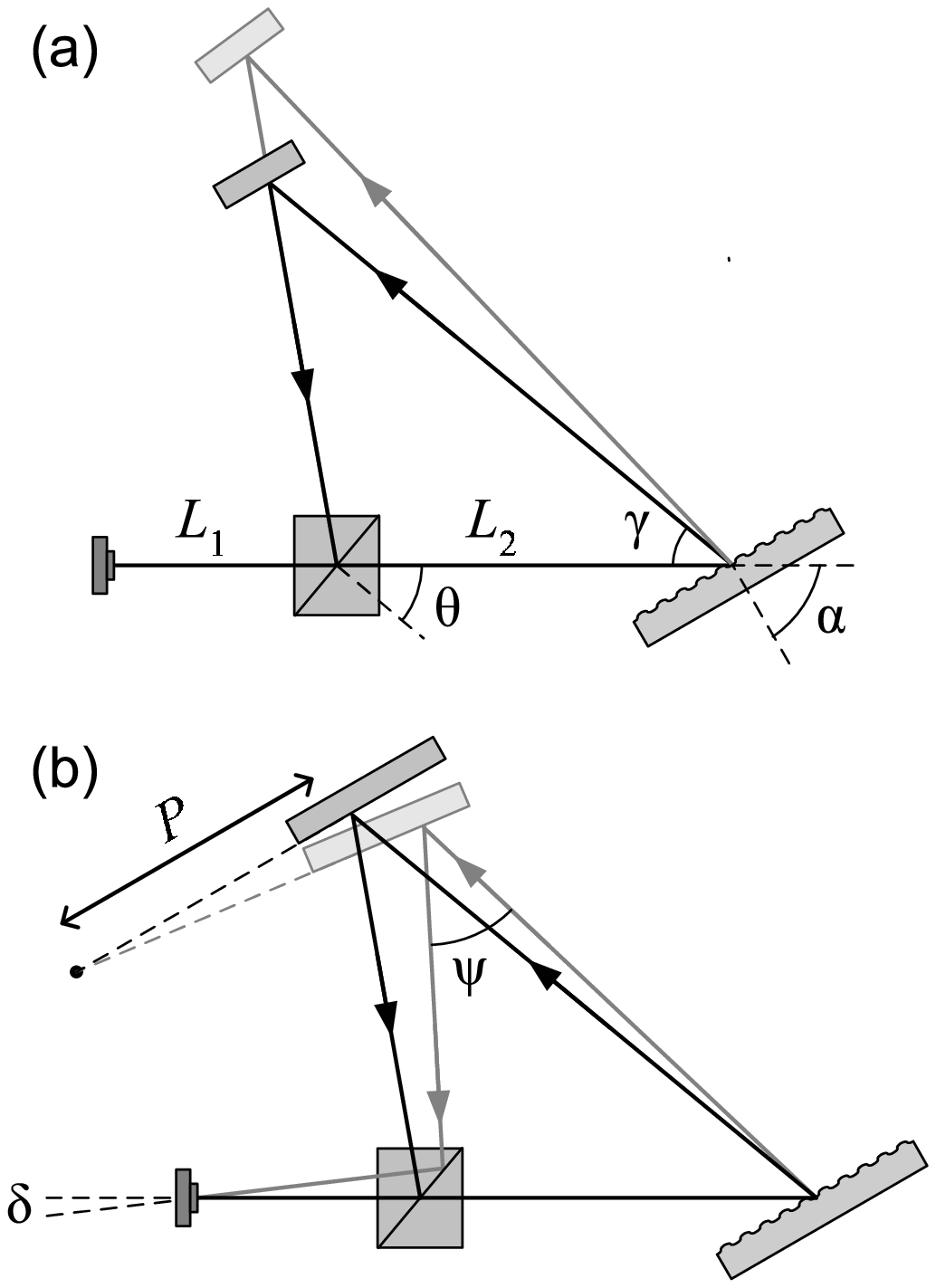}

\caption{Tuning the laser.  The four parameters $L_1$, $L_2$,
$\theta$, and $\alpha$ which, along with the nominal laser
wavelength and the grating line spacing, define the cavity, are
shown in (a). The angle $\gamma$ is determined from Eq.\
\ref{eq:gamma}. Shortening the cavity at higher $\gamma$ can be
done by letting the beam return to the laser diode at a small
angle $\delta$ relative to the outgoing beam, as shown in (b).
\label{fig:tuning}}
\end{figure}

Using the simplest model of our laser, in which it is assumed that
the light fed back to the laser exactly retraces the path of the
outgoing beam, it would appear that it is impossible to scan our
laser in a way which satisfies Eq.\ \ref{eq:modehopcondition}. In
Fig.\ \ref{fig:tuning}(a) it can be seen that by rotating and
translating the upper mirror, it is possible to increase the
angle between the beam incident upon the grating and the
diffracted beam (this angle is denoted as $\gamma$ in the
figure). When this is done the length of the external cavity {\em
increases}, while the wavelength of light diffracted with
increasing $\gamma$ {\em decreases}, causing the feed-back
wavelength and the cavity length to scan in opposite directions.
Using this model we would predict that our laser should only be
able to scan about 100 MHz before it became favorable to hop to
another mode.  The fact that we have been able to scan much
further without mode hops indicates that this model is incomplete.

In our current model, the returning beam is allowed to be at a
slight angle $\delta$ relative to the beam exiting the diode (see
Fig.\ \ref{fig:tuning}(b)).  Since the collimated laser beam has a
finite width, the beam contains a spread of wavevectors.  By
measuring the Gaussian radius $w_0$ of the collimated beam we can
calculate the $1/e^2$ full angular divergence of the beam: $\phi
= 2 \lambda / \pi w_0$.  In our model we assume that a beam
returning with a $ \delta $ which is considerably less than $\phi
/ 2$ will couple back to the diode nearly as well as a beam going
straight back with $\delta = 0$.

To calculate the tuning range using this model, we first used
simple trigonometry to calculate the round trip path length for a
cavity.  The optical path length from the laser diode to the
rejection port polarizing beamsplitter of the isolator (labeled as
$L_1$ in Fig.\ \ref{fig:tuning}), and the optical path length from
the beamsplitter to the grating (labeled $L_2$ in Fig.\
\ref{fig:tuning}) were assumed to be known quantities.  In
addition, the angles between the incident beam and the normal
vectors of the beamsplitter and grating (labeled $\theta$ and
$\alpha$, respectively), the spacing between lines on the grating
$d$, and the nominal ($\delta=0$) wavelength of the laser
$\lambda_0$ were assumed to be known. These six quantities define
the configuration of a particular laser.

The calculated path length, $S$, is a function of the known
parameters as well as the feedback angle $\delta$ and the angle
between the incident and diffracted beam at the grating
$\gamma$.  Using the grating equation and assuming that the first
diffraction order from the grating is the one fed back to the
diode, the angle $\gamma$ can be solved for in terms of the
wavelength of the laser:
\begin{equation}
    \gamma = \alpha - \arcsin \left( \lambda/d - \sin(\alpha) \right)\label{eq:gamma}
\end{equation}
Substituting this relation for $\gamma$ in our expression for $S$
resulted in an equation for the round trip path length which is
only a function of known quantities, the angle $\delta$, and the
wavelength $\lambda = \lambda_0 + \Delta \lambda$.

Next we substituted our expression for $S$ into Eq.\
\ref{eq:modehopcondition} to generate an equation which relates
the wavelength shift $\Delta \lambda$ to $\delta$ under the
condition that the mirror is moved and rotated in the manner which
satisfies the criterion for mode-hop free tuning.  This somewhat
complicated equation can be solved numerically to find $\Delta
\lambda$ for a given $\delta$.  In order to generate an
analytical solution, we first linearized this equation in
$\delta$ and were then able to solve the resulting first-order
equation for $\Delta \lambda$. Then, because the tuning range of
a diode laser is more often discussed in terms of frequency than
wavelength, we converted this to an equation for the frequency
detuning, $\Delta f$, with the first order relationship $\Delta f
= \Delta \lambda c / \lambda_0^2$. Finally, we calculated the
tuning range of the laser by assuming that allowed values of
$\delta$ ranged from $-\phi /2$ to $\phi /2$.

The final result of this calculation is an equation for detuning
which is just the maximum allowed range of $\delta$ times a
constant:
\begin{equation}
    \Delta f =  Q \delta \label{eq:tuningrange}
\end{equation}
The tuning range of the laser can then be approximated by taking
the difference between $\Delta f$ calculated at $\delta = \phi /
2 $ and at $\delta = -\phi / 2$, which gives a tuning range of $Q$
times the full-angle Gaussian divergence $\phi$.  The
proportionality constant $Q$ is given by the following expression.
\begin{equation}
    Q = \frac{c \left(L_1 A + L_2 B \right)}{\lambda_0 \left( S_0 +  L_2 \lambda_0 C / d \right)
    } \label{eq:Q}
\end{equation}
Here $c$ is the speed of light and $\lambda_0$ is the nominal
wavelength of the laser.  The unitless parameters $A$, $B$, and
$C$ are given by
\begin{equation}
    A \equiv \frac{ 1 + \cos \psi_0 }{
    \sin \psi_0 }
\end{equation}
\begin{equation}
    B \equiv \frac{ \sin \gamma_0 }{ 1 - \cos \psi_0}
\end{equation}
and
\begin{equation}
    C \equiv \frac{ \sin  2 \theta  }{ \cos \left(
    \alpha - \gamma_0 \right) \left[ 1 - \cos \psi_0 \right]}
\end{equation}
where $\gamma_0$ is the angle between the incident and diffracted
beam at the grating when $\delta = 0$. This angle can be measured
physically for a particular laser or can be calculated by setting
$\lambda = \lambda_0$ in Eq.\ \ref{eq:gamma}.  The angle $\psi_0 =
2 \theta - \gamma_0 $ is the angle between the incident and
reflected beams at the tuning mirror (see Fig.\
\ref{fig:tuning}(b)) when $\delta = 0$.

The $S_0$ term in Eq.\ \ref{eq:Q} is the round-trip optical path
length of the cavity when $\delta = 0$, given by the relation
\begin{equation}
    S_0 = 2 L_1 + \left( 1 + \frac{\sin  2 \theta
         + \sin  \gamma_0 }{\sin \psi_0 } \right) L_2
\end{equation}

Using the parameters of the laser which we tested, the first-order
model predicts a mode-hop free tuning range of $\simeq 4 \times
10^{11}$ Hz (or about 0.6 nm), well in excess of the measured
range, implying that we have not realized the maximum possible
tuning range for our configuration. This first-order tuning range
estimate agrees with the predicted tuning range determined from a
full numerical solution to better than 0.1\%.

Although ``ideal tuning'' in this configuration is not achieved by
simply pivoting the mirror about a fixed axis, in many cases this
simple method is close enough to the ideal geometry that the full
tuning range predicted by Eq.\ \ref{eq:tuningrange} can be
achieved.  For example, a complete numerical model of our current
laser configuration reveals that if the correct pivot point is
chosen, tuning by simply pivoting the mirror results in a change
of $m$ in Eq.\ \ref{eq:nnodes} by only $0.08$ over the entire
tuning range predicted by Eq.\ \ref{eq:tuningrange}.  The precise
location of this pivot point, however, is very important. The
numerical model of our laser shows that changing the pivot point
by $ \pm 1$ mm reduces the expected scan range by more than an
order of magnitude.

To calculate the ideal pivot lever arm $P$ (see Fig.\
\ref{fig:tuning}(b)) for a maximum mode-hop free tuning range, we
first used Eqs.\ \ref{eq:gamma} and \ref{eq:tuningrange} to find
$\gamma$ as a function of $\delta$.  This relation was reduced to
lowest order in $\delta$ to produce the equation $\gamma =
\gamma_0 + G \delta$ where $G$ is given by
\begin{equation}
    G  = \frac{Q \lambda_0^2}{c d \cos \left( \alpha - \gamma_0 \right) }
\end{equation}
We were then able to find the angle $\psi = $ $2 \theta_0 - \gamma
+ \delta \simeq $ $2 \theta_0 - \gamma_0 + (1-G) \delta$ in Fig.\
\ref{fig:tuning}(b), as well as the location of the point in space
where the beam reflects off of the mirror, as a function of known
quantities and $\delta$.  From this, and using the law of
reflection, we were able to write down an equation for the line
which follows the surface of the mirror (indicated by the dotted
lines near the top of Fig.\ \ref{fig:tuning}(b)) as a function
$\delta$ in slope-intercept form. We then set both sides of the
equation for a line at finite $\delta$ equal to the equation for
the $\delta=0$ line.  After linearizing this relation we solved
for the point at which these two lines cross.  The distance from
this point to the place at which the beam strikes the mirror when
$\delta=0$ is given by the relation
\begin{equation}
    P =  \frac{ 2 \sin \left( \theta + \gamma_0/2 \right) \left( L_1 D + L_2 E \right) }{ \left( 1 + G \right) \sin^2
    \psi_0 } \label{eq:pivot}
\end{equation}
where the unitless parameters $D$ and $E$ are given by
\begin{equation}
    D   \equiv
     \frac{ \sin \psi_0 }{ \sin \left( 2 \theta + \gamma_0 \right)}
     \left[ \cos  2 \theta  + \cos
     \gamma_0   \right]
\end{equation}
and
\begin{equation}
    E \equiv \frac{ \cos 2 \theta + \cos \gamma_0 }{ \sin \left( 2
    \theta + \gamma_0 \right)} \left[ \sin \gamma_0 - G \sin 2 \theta
    \right]
\end{equation}

\section{Conclusion}

In conclusion, we have demonstrated a new scheme for external
cavity diode laser stabilization which combines the higher output
of the Littrow scheme with the stable output pointing of the
Littman-Metcalf scheme.  We have measured the spectral properties
of a prototype laser, and found them to be comparable to typical
Littrow lasers.  We have measured a mode-hop free scan range of
several GHz when tuning the laser by pivoting the mirror about a
fixed axis, and we have developed a model which predicts that
much larger tuning ranges are possible.

\section*{Acknowledgments}

This work was supported in part by grants from the Research
Corporation and from the National Science Foundation (Grant No.
PHY-9985027).


\begin{thebibliography}{}
\bibitem{circulaser}
Circu-laser diodes have a cylindrical lens mounted in the 9 mm
package to produce a beam with symmetric divergence.
\end{thebibliography}
\end{document}